\documentclass{article}
\usepackage{lineno}
\usepackage{geometry}
\usepackage[version=4]{mhchem}
\usepackage{gensymb}
\usepackage{lineno}
\usepackage{amsmath,amssymb}
\usepackage{wasysym}
\usepackage{float}
\usepackage{longtable}
\usepackage{times}
\usepackage{multicol}

\usepackage{multirow}
\usepackage{graphicx}
\usepackage{natbib}

\usepackage{placeins} 
\graphicspath{{./}{figures/}{figures/revision_220920/}}

\newcommand{\aap}{A\&A}                   
\newcommand{\apjs}{ApJS}                  
\newcommand{\apjl}{ApJ}                   

\newcommand{\m}{\rm\thinspace m}

\newcommand{\s}{\rm\thinspace s}

\newcommand{\K}{\rm\thinspace K}

\newcommand{\g}{\rm\thinspace g}
\begin{document}

\title{CO$_2$ ocean bistability on terrestrial exoplanets}


\author{R.J. Graham\footnote{robert.graham@physics.ox.ac.uk}, Tim Lichtenberg, \& Raymond T. Pierrehumbert}

\maketitle

\begin{center}
    \small
   Atmospheric, Oceanic and Planetary Physics, Department of Physics\\
   University of Oxford, United Kingdom 
\end{center}


\begin{abstract}
Cycling of carbon dioxide between the atmosphere and interior of rocky planets can stabilize global climate and enable planetary surface temperatures above freezing over geologic time. However, variations in global carbon budget and unstable feedback cycles between planetary sub-systems may destabilize the climate of rocky exoplanets toward regimes unknown in the Solar System.
Here, we perform clear-sky atmospheric radiative transfer and surface weathering simulations to probe the stability of climate equilibria for rocky, ocean-bearing exoplanets at instellations relevant for planetary systems in the outer regions of the circumstellar habitable zone.
Our simulations suggest that planets orbiting G- and F-type stars (but not M-type stars) may display bistability between an Earth-like climate state with efficient carbon sequestration and an alternative stable climate equilibrium where \ce{CO2} condenses at the surface and forms a blanket of either clathrate hydrate or liquid \ce{CO2}. At increasing instellation and with ineffective weathering, the latter state oscillates between cool, surface \ce{CO2}-condensing and hot, non-condensing climates.
\ce{CO2} bistable climates may emerge early in planetary history and remain stable for billions of years. The carbon dioxide-condensing climates follow an opposite trend in $p$\ce{CO2} versus instellation compared to the weathering-stabilized planet population, suggesting the possibility of observational discrimination between these distinct climate categories.
\end{abstract}

\section*{Key Points}
\begin{itemize}
\item At low instellations within the circumstellar habitable zone, rocky planets with H$_2$O oceans may also develop oceans of condensed CO$_2$.

\item CO$_2$-ocean-bearing climate states are bistable with more traditional Earth-like climates where only water condenses at the planetary surface

\item CO$_2$ oceans may occur even in the presence of the negative feedback on planetary climate provided by silicate weathering.
\end{itemize}

\section*{Plain Language Summary}

On Earth, water is the only molecule that occurs in large quantities as both vapor and condensed phases, but that may not be the case for every planet. We simulate terrestrial exoplanets that receive less light from their stars than the Earth does. We find that \ce{CO2} may build up to such high levels that it condenses out onto the planetary surface, allowing for oceans of liquid \ce{CO2} and/or sheets of \ce{CO2} clathrate to accumulate. Depending on factors like \ce{CO2} outgassing rate, the kinds of silicate rocks present at the planetary surface, and the level of irradiation received by a given planet from its parent star, such oceans can remain stable for geologic-scale time periods, or a planet may oscillate back and forth between an Earth-like, non-\ce{CO2}-condensing state and an exotic \ce{CO2}-condensing state. Planets with liquid (or solid) \ce{CO2} at their surface can have a profoundly different evolution than Earth, with important implications for their potential to host life.

\twocolumn
\section{Introduction} \label{sec:intro}
Earth's surface is dominated by a liquid water ocean in direct contact with the lithosphere. This state of affairs seems to be crucial for Earth's long-term climate stability and habitability, with the carbonate-silicate cycle modulating and stabilizing the planet's atmospheric \ce{CO2} inventory through a set of water-rock chemical reactions taking place on the continents and seafloor \citep{Walker-Hays-Kasting-1981:negative, berner1983carbonate, coogan2013evidence,penman2020silicate}. The carbonate-silicate cycle acts as a thermostat when \ce{CO2} is a net greenhouse gas; however, when \ce{CO2} builds to high enough levels, it increases planetary Rayleigh scattering and behaves as a coolant, which can convert the carbonate-silicate cycle into a destabilizing positive feedback. This suggests that otherwise Earth-like planets with large enough carbon inventories might be able to support climate configurations with high enough $p$\ce{CO2} for either a hot, supercritical, Venus-like atmosphere at high instellations or an exotic, subcritical atmosphere with surface liquid \ce{CO2} condensation coexisting with a liquid water ocean at low instellations. Planets of the latter variety might be difficult to remotely distinguish from more traditionally ``Earth-like'' planets lacking surface \ce{CO2} condensation at equivalent orbits, but the geochemistry and potential habitability of these worlds would be radically different, even with a temperate surface climate. Most previous examinations of surface \ce{CO2} condensation on terrestrial (exo)planets have focused on cold, glaciated climates where \ce{CO2} would only condense as a solid \citep{turbet2017co,kadoya2019outer,bonati2021influence}; waterworlds with high pressure ice mantles \citep{ramirez2018ice,marounina2020internal}; or the potential for CO$_2$ condensation on Mars in the deep past \citep{kasting1991co2,forget20133d,soto2015martian}. In this study, we focus on surface CO$_2$ condensation on rocky exoplanets with temperate climates in different end-member weathering regimes that inform the anticipated diversity of potentially habitable planets \citep[][]{kasting93, wordsworth2010gliese,von2013atmospheric}.

From an astronomical perspective, carbon compounds are strongly depleted on the terrestrial planets of the Solar System relative to the nominal values in the Sun or the interstellar medium \citep{2021PhR...893....1O} as a result of processes operating in the protoplanetary disk \citep{krijt2020,Li+21} and on planetesimals \citep{Hirschmann21,2021ApJ...913L..20L}. In addition, volatile partitioning into metal and melt phases can redistribute major carbon and hydrogen carriers between core, mantle, and atmospheric reservoirs and partly decouple the initially accreted volatile reservoir \citep{2020GeCoA.280..281G,Fischer2020} from the atmospheric composition of rocky exoplanets. A recent example is provided by the outer TRAPPIST-1 planets, for which mass-radius contraints suggest volatile mass fractions on the order of several weight per cent \citep{2021PSJ.....2....1A}. On a statistical level, the larger sub-Neptune cluster of the Kepler radius valley suggests that at least a fraction of systems accrete substantial volatile mass budgets during their formation \citep{2019PNAS..116.9723Z,2020A&A...643L...1V}. The anticipated variation in carbon abundance suggests that the majority of rocky exoplanets may exhibit diverse climate regimes, for which the thermodynamic limits to maintain clement surface states are poorly understood \citep{2022arXiv220310023L}. Future exoplanet surveys that will aim to probe the atmospheres of temperate exoplanets to test the range of climate diversity \citep{HABEX_StudyReport2019,LIFE2021a,2021AJ....161..150C} will rely on physically motivated theories to interpret their findings. 

Here, we study the interplay between silicate weathering and \ce{CO2} pressure variations to probe the limits of clement climates on terrestrial exoplanets. We use 1-D, two-stream radiative transfer and carbon cycle simulations to investigate the behavior of climates with high partial pressure of \ce{CO2} ($p$\ce{CO2}) and low irradiation from the central star ($S_{\rm eff}$) without (sections \ref{subsec:no_weathering}, \ref{subsec:limit_cycling}) and with (section \ref{subsec:bistability}) weathering feedbacks. Our simulations suggest that terrestrial planets at low instellations in the classical circumstellar habitable zone \citep[HZ;][]{kasting93} may emerge from their accretionary period directly into stable climate states with long-lasting periods of liquid \ce{CO2} surface condensation, even in cases where they are not initially globally glaciated. 

\section{Methods}\label{sec:modeling}
In this study we combine global-mean, clear-sky climate and silicate weathering calculations to examine the interplay between radiative and geochemical feedbacks on ocean-bearing, high-$p$\ce{CO2} planets in the outer reaches of the classical habitable zone. Here we briefly outline the procedure we follow and the models we use for the radiative calculations and the weathering calculations.

\label{appsec:methods}
\subsection{Radiative transfer}
\label{subsec:rad}
We carry out radiative transfer calculations using the \textsc{socrates} code \citep{edwards1996studies}, solving the plane-parallel, two-stream approximated radiative transfer equation with scattering \citep[see the extensive description in][though note that the implementation in that paper does not include scattering]{lichtenberg2021vertically}. Opacity coefficients are tabulated and derived from the HITRAN database, making use of the line-by-line and collision-induced absorption coefficients for H$_2$O \citep{HITRAN2016}, \ce{CO2} \citep{HITRAN2016,gruszka1997roto}, N$_2$ \citep{HITRAN2016,karman2015quantum}, and the H$_2$O continuum \citep{mlawer2012development}. We note that the CO$_2$ continuum spectrum is uncertain at high temperatures and pressures, which introduces a potentially significant source of error into our calculations \citep[e.g.][]{Halevy09,wordsworth2010infrared}.

Rayleigh scattering cross-sections for \ce{CO2} and N$_2$ are calculated following \citet{vardavas1984solar}
\begin{linenomath}
\begin{align}
\sigma_{R,i} &= \frac{0.2756}{\mu_i}\times\frac{6+3\Delta}{\lambda^4(6-7\Delta)}[A(1 + \frac{B}{\lambda^2})]^2,
\end{align}
\end{linenomath}
where the subscript $i$ iterates over the species present, $\sigma_{R,i}$ [m$^2$ kg$^{-1}$] is the Rayleigh scattering cross-section, $\mu_i$ is the molar mass of species $i$ [kg mol$^{-1}$], $\lambda$ [$\mu\rm m$] is wavelength, coefficients $A$ and $B$ are taken from \citet{cox2015allen}, $\Delta$ is the depolarization factor, and the numerical values we use are given in Table \ref{tab:rayleigh}. For H$_2$O, as far as we are aware, values for the coefficients $A$ and $B$ have not been published at the relevant wavelengths. For this reason, and because H$_2$O is a minor constituent in the atmospheres we simulate, we use a simple $\lambda^{-4}$ scaling to calculate H$_2$O's Rayleigh scattering cross-section,
\begin{linenomath}
\begin{align}
    \sigma_{R,\rm H_2O} = \sigma_{R,0} \frac{\lambda_0^4}{\lambda^4},
\end{align}
\end{linenomath}
where $\sigma_{R,0}=9.32\times10^{-7}$ m$^2$ kg$^{-1}$ \citep{Pierrehumbert:2010-book} and $\lambda_0$ = 1 $\mu$m. Here we take the opportunity to note that some previous papers \citep{Kopparapu:2013, pluriel2019modeling}, have erroneously used H$_2$O Rayleigh scattering coefficients calculated using a depolarization ratio that was drawn from a study \citep{marshall1990raman} of the scattering properties of \textit{liquid} H$_2$O, not water vapor. The total Rayleigh scattering cross-section is calculated by summing the cross-sections of the individual species, weighted by volumetric mixing ratio,
\begin{linenomath}
\begin{align}
    \sigma_{R,\rm tot} &= \sum_i x_i\sigma_{R,i},
\end{align}
\end{linenomath}
where $x_i$ represents the volumetric mixing ratio of a given species ($\sum_i x_i=1$). \textsc{socrates} does not allow vertically-varying Rayleigh scattering coefficients, so we take the mixing ratios at the surface to calculate the total scattering cross-sections. 

The primary stellar spectrum we use in the presented calculations is based on measurements of the Sun's spectral irradiance \citep{Kurucz1995}, and thus represents irradiation from a G2V star. To show how the climate behaviors we identify depend on stellar type and age, we also present sets of simulations using spectra from AD Leonis, an M3.5V star \citep{Segura:2005}, and Sigma Boötis, an F2V star \citep{segura03}. The different spectra result in different planetary albedo values for a given atmospheric composition and climate. We also tested the effects of a change in the solar spectrum with time: at 4.5 and 3.8 Ga before present \citep{claire2012evolution} our simulations produced results that differed negligibly from the fiducial, modern case.

\begin{figure*}[htb]
    \centering
    \makebox[\textwidth][c]{\includegraphics[width=400pt,keepaspectratio]{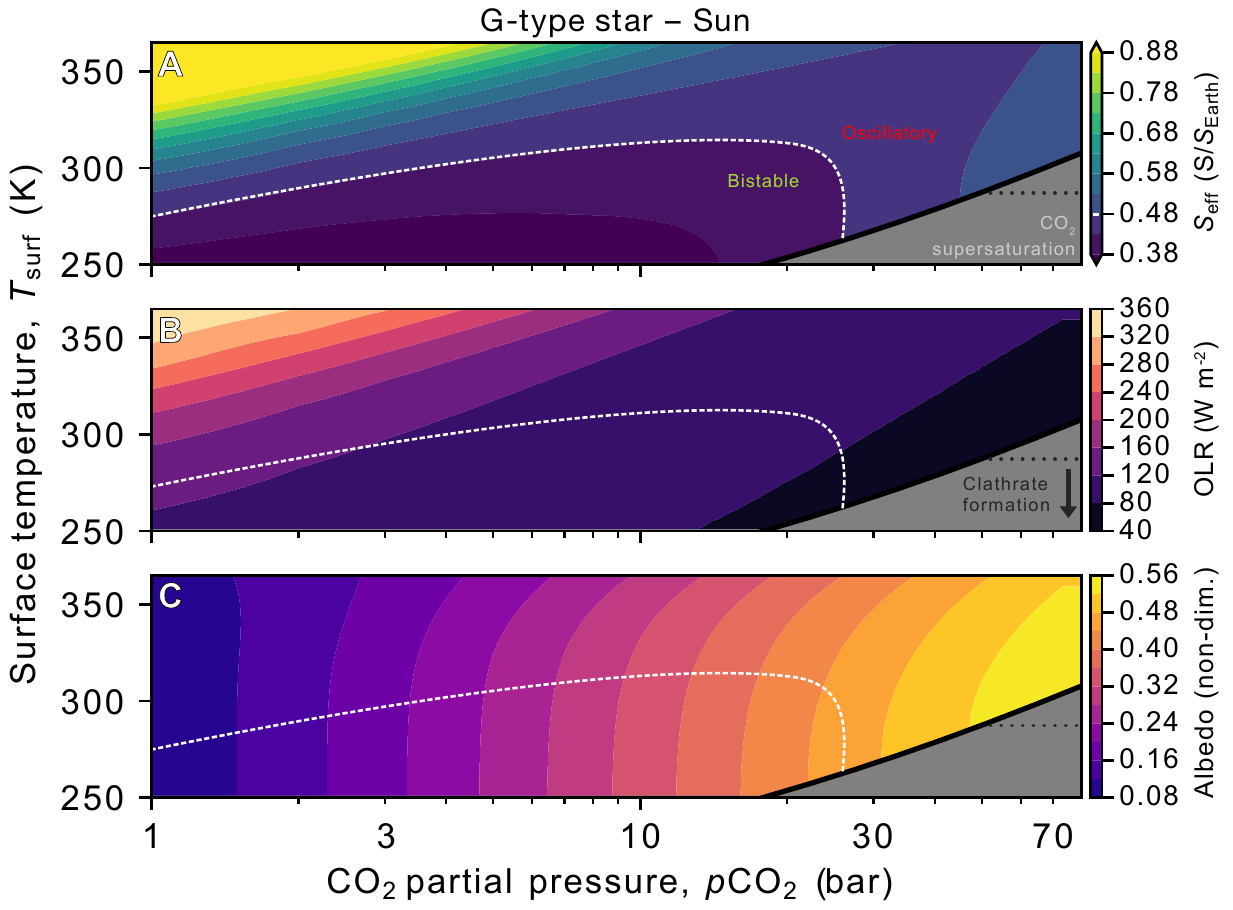}}
    \caption{Energetic properties of terrestrial planetary climates at low instellation as a function of surface temperature $T_\mathrm{surf}$ ($y$-axes) and \ce{CO2} partial pressure ($x$-axes). In each panel, the black line represents the \ce{CO2} saturation vapor pressure curve; climates with \ce{CO2} greater than that of the vapor pressure curve at a given temperature are super-saturated at their surface (grey region in all three panels). The white dotted line in each panel represents the approximate contour location of the $S_{\rm eff}$ value separating climates that can support \ce{CO2} bistability vs. those that may limit cycle between condensing and non-condensing states (see Sections \ref{subsec:limit_cycling} and \ref{subsec:bistability} for discussion of those climate types). The black dotted line in each panel marks the temperature below which \ce{CO2} clathrate hydrates may be stable. In panel A, the contours represent the top-of-atmosphere stellar irradiation required to maintain a climate with a given surface temperature and $p$\ce{CO2}, normalized by the instellation received at Earth's orbit, i.e. $S / S_{\rm Earth} = S_{\rm eff}$. Panel B shows the global-mean outgoing longwave radiation (OLR). Panel C shows the global-mean planetary albedo.}
    \label{fig:Seff_OLR_albedo}
\end{figure*}

We apply a generalized pseudoadiabat formula \citep{graham2021multi} to calculate atmospheric temperature-pressure profiles with a variety of compositions and surface boundary conditions. This pseudoadiabat formula incorporates the fraction of retained condensate as a freely tunable parameter which can significantly impact the specific heat capacity and lapse rate in atmospheres with non-dilute condensable species. In our calculations, we assume that all condensate is instantaneously rained out upon condensation. We also assume H$_2$O saturation in the main set of simulations discussed in this paper. This formula is useful because it allows us to self-consistently calculate atmospheric profiles with any combination of condensable (e.g. H$_2$O, \ce{CO2}) and non-condensable (e.g. N$_2$) gases, though the atmospheres we focus on in this paper are simply \ce{CO2}+H$_2$O (sensitivity tests with up to 10 bar of N$_2$ produced qualitatively identical results, as expected). Throughout this paper, unless otherwise noted, atmospheres are taken to have isothermal stratospheres with $T_{\rm strat} = 150$ K. This is comparable to the stratospheric temperatures hypothesized for high-\ce{CO2} planets in the outer reaches of the classical circumstellar habitable zone \citep{Kopparapu:2013}, and time-stepped radiative-convective calculations in high-pCO$_2$ atmospheres have recovered a stratospheric temperature of 150 K \citep{wordsworth2013water}. Condensation of H$_2$O and \ce{CO2} is assumed to cease within the stratosphere, so the mixing ratios remain constant at pressure levels below the highest pressure level (lowest altitude level) with a temperature of 150 K. Assuming a constant stratospheric temperature considerably simplifies our climate calculations (described further below), at the expense of neglecting the feedback between instellation and stratospheric temperature, which can change a planet's outgoing longwave radiation and thus its surface temperature. Sensitivity tests carried out with an increased stratospheric temperature of up to 200 K demonstrate that the effect on climate is minor, with no qualitative changes to our results.

For rapid simulation of a wide range of surface conditions, we take an ``inverse climate modeling'' approach \citep{kasting1991co2}. This entails choosing a $p$\ce{CO2} and a surface temperature (which in turn specifies $p$H$_2$O by the Clausius-Clapeyron relation), using those values as boundary conditions to integrate the pseudoadiabat from the surface up to the 150 K isothermal stratosphere, and running the radiative transfer code to get the OLR and albedo for that specific atmospheric temperature/pressure/composition combination. With those values, we can calculate the instellation necessary to maintain global-mean energetic balance between incoming and outgoing radiation,
\begin{linenomath}
\begin{align}
\label{eqn:energetic_equilibrium}(1-\alpha(T,\mathrm{pCO}_2)\frac{S}{4} &= F_{\rm out}(T,\rm pCO_2),\\
S_{\rm eff}(T, \mathrm{pCO}_2) = \frac{S}{S_{\rm Earth}} &= \frac{4F_{\rm out}(T, \rm pCO_2)}{S_{\rm Earth}(1-\alpha(T, \rm pCO_2))},
\end{align}
\end{linenomath}
where $S$ is top-of-atmosphere instellation, $S_{\rm Earth} = 1368$ W m$^{-2}$ is Earth's present-day instellation, $S_{\rm eff}$ is the fraction of present-day Earth's instellation (e.g. $S_{\rm eff} = 0.3$ is equivalent to 30$\%$ of present-day Earth instellation), $F_{\rm out}$ is the OLR, and $\alpha$ is the global-mean planetary albedo, with the cosine of the stellar zenith angle assumed to be the instellation-weighted global mean of $\frac{2}{3}$ in all calculations \citep{cronin2014choice}. We also set the surface albedo to 0.0, similar to the albedo of a cloudless sea surface, which would be 2--4$\%$ at the chosen stellar zenith angle \citep{li2006ocean}. Low-lying marine stratocumulus clouds that would increase the near-surface albedo to levels above that of the sea surface are expected to dissipate at \ce{CO2} levels far lower than 1 bar, the lowest $p$\ce{CO2} evaluated in this study, due to the inhibition of cloud-top radiative cooling and subsequent shutdown of cloud-sustaining lower-tropospheric convection \cite[e.g.][]{schneider2019possible}, though of course other processes could still cause low-lying clouds or hazes that would affect near-surface albedo on these planets. 

We carried out a grid of these inverse climate calculations for $p$\ce{CO2} levels ranging from 1 bar to 73 bar with increments of 1 bar and for $T_{\rm surf}$ from 250 K to 365 K with increments of 5 K. Fig.~\ref{fig:Seff_OLR_albedo} shows equilibrium $S_{\rm eff}$ (panel A), OLR (panel B), and albedo (panel C), all as functions of $T_{\rm surf}$ and $p$\ce{CO2}. Linear interpolation of OLR and albedo to temperatures and $p$\ce{CO2} levels between the climate grid points allows for fast climate calculations to examine a wide variety of scenarios.

\subsection{Carbon cycling}
\label{subsec:mac_model}
To examine how the carbon cycle might operate on abiotic terrestrial planets under very high $p$\ce{CO2} conditions, we apply an idealized global-mean weathering formulation \citep{graham2020thermodynamic} based on work that accounts for the impact of clay precipitation on weathering solute concentrations \citep{maher2014hydrologic} on the global-mean weathering flux,
\begin{linenomath}
\begin{align}
    W &= \label{eqn:mac_weathering}\gamma\frac{\alpha}{[k_{\rm eff}]^{-1}+mAt_{\rm s}+\alpha[qC_{\rm eq}]^{-1}},
\end{align}
\end{linenomath}
where $W$ [mol m$^{-2}$ yr$^{-1}$] is the global-mean weathering flux, i.e. the total number of divalent cations (which react with oceanic carbon to form carbonate minerals, ultimately removing \ce{CO2} from the atmosphere) delivered to the ocean from the land and/or seafloor in a year, divided by the surface area of the planet; $\gamma$ is the fraction of planetary surface that is weatherable; $\alpha$ is a parameter that captures the effects of various weathering zone properties like characteristic water flow length scale, porosity, ratio of mineral mass to fluid volume, and the mass fraction of minerals in the weathering zone that are weatherable; $k_{\rm eff}=k_{\rm eff, ref}\exp{\left(\frac{T_{\rm surf}-T_{\rm surf,ref}}{T_e}\right)}\left(\frac{p\text{CO}_2}{p\text{CO}_{\rm 2,ref}}\right)^\beta$ [mol m$^{-2}$ yr$^{-1}$] is the effective kinetic weathering rate, i.e. the weathering rate in the absence of chemical equilibration with clay precipitation \citep{Walker-Hays-Kasting-1981:negative}; $m$ [kg mol$^{-1}$] is the average molar mass of minerals being weathered; $A$ [m$^{2}$ kg$^{-1}$] is the average specific surface area of the minerals being weathered; $t_s$ [yr] is the mean age of the material being weathered; $q=q_{\rm ref}(1+\epsilon(T_{\rm surf} - T_{\rm surf, ref}))$ [m yr$^{-1}$] is the volume-weighted global-mean flux of water through the planet's weathering zones (in this study we apply a linear temperature dependence to runoff, based on the behavior of surface H$_2$O precipitation on Earth, but we note that the functional form for $q$ could be very different when modeling seafloor weathering); and $C_{\rm eq}=\Lambda(\text{pCO}_2)^n$ [mol m$^{-3}$] is the maximum concentration of divalent cations in the water passing through weathering zones, as determined by chemical equilibrium between silicate dissolution and clay precipitation. The values of all constants are given in Table \ref{tab:weathering_values}. A more thorough set of weathering calculations would carefully account for the lithology of minerals being weathered \citep{hakim2021lithologic} and the differences between continental and seafloor weathering \citep{hayworth2020waterworlds, hakim2021lithologic}, but our calculations are meant to be illustrative of the qualitative behavior of the carbon cycle under non-terrestrial conditions, so we restrict our simulations to the simplified approach in our equation \ref{eqn:mac_weathering}.

Assuming the presence of weatherable silicates on a planetary surface, the global weathering flux of divalent cations into the ocean, which is equal to the global \ce{CO2} consumption flux at equilibrium and in the absence of \ce{CO2} surface condensation, is dependent on both background \ce{CO2} and $T_{\rm surf}$. An increase to $p$\ce{CO2} or $T_{\rm surf}$ leads to an increase in the weathering rate, with the change mediated by either changes to the kinetics of silicate dissolution or changes to global runoff flux, depending on which term in equation \ref{eqn:mac_weathering} is dominant. This climate-dependence of \ce{CO2} consumption means that silicate weathering can act as a stabilizing negative feedback on planetary climate \citep{Walker-Hays-Kasting-1981:negative}. If the \ce{CO2} outgassing flux from volcanoes and other sources ($V$) is greater than the \ce{CO2} consumption flux from weathering, i.e. if $V>W$, and if there is no \ce{CO2} surface condensation, \ce{CO2} will accumulate in the atmosphere, which, under Earth-like circumstances, tends to warm the planet. Higher $p$\ce{CO2} and higher $T_{\rm surf}$ both lead to larger $W$, driving the \ce{CO2} consumption rate closer and closer to $V$ until \ce{CO2} consumption is equal to \ce{CO2} production and the atmospheric \ce{CO2} inventory stabilizes. The same process in reverse acts to cool the planet and equilibrate the carbon cycle in cases where $W>V$. So, at least in the cases just discussed, climate on planets with silicate weathering will tend to find an equilibrium $T_{\rm surf}$ and $p$\ce{CO2} determined by the balance between silicate weathering and \ce{CO2} outgassing, which can be stated simply as
\begin{linenomath}
\begin{align}
    V&=W\label{eqn:carbon_balance},
\end{align}
\end{linenomath}
where $V$ [mol m$^{-2}$ yr$^{-1}$] is an assumed \ce{CO2} outgassing flux and $W$ is the weathering flux as defined in equation \ref{eqn:mac_weathering}. The intersection points in $T_{\rm surf}$--$p$\ce{CO2} space of the nullclines given by equations \ref{eqn:energetic_equilibrium} and \ref{eqn:carbon_balance} are climate states in equilibrium with respect to both energy and carbon fluxes.

The weathering parameterization represented by equations \ref{eqn:mac_weathering} and \ref{eqn:carbon_balance} implies the assumption of an Earth-like tectonic regime where the resurfacing of fresh silicates occurs rapidly enough to maintain a weathering flux in balance with CO$_2$ outgassing. This need not be the case: for example, if a planet is in a ``sluggish lid'', ``episodic lid'', or ``stagnant lid'' tectonic regime \citep[e.g.][]{valencia:2007p1901,korenaga2010likelihood,kite:2009p2923,foley2015role,lenardic2018diversity,2022arXiv220310023L}, resurfacing may not be fast enough for weathering to keep up with the outgassing rate, leading to a global depletion of weatherable materials called ``supply limitation'' \citep{west2005tectonic} or ``transport limitation'' \citep{kump2000chemical}, which in turn allows for volcanic CO$_2$ accumulation. Further, even with rapid tectonic resurfacing, the particular climate and/or arrangement of land on a given planet may not allow for high enough weathering fluxes to match outgassing rates, as we will go on to demonstrate. For these cases, it is important to note that surface \ce{CO2} condensation can act as another major sink of atmospheric \ce{CO2} \citep{kasting1991co2,wordsworth2010gliese,von2013atmospheric, turbet2017co,kadoya2019outer,bonati2021influence}. As a result, under \ce{CO2}-condensing conditions, it is possible for the carbon cycle to reach equilibrium even when outgassing does not equal weathering, as condensation can make up the difference,
\begin{linenomath}
\begin{align}
\label{eqn:cond_flux}
    F_{\rm cond} &= V - W,
\end{align}
\end{linenomath}
where $F_{\rm cond}$ [mol m$^{-2}$ yr$^{-1}$] is the flux of \ce{CO2} condensing out onto the surface from the atmosphere. 

\section{Results}\label{sec:results}

\subsection{Climate hysteresis from temperature-dependent instellation absorption of \ce{H2O}}
\label{subsec:no_weathering}
In climate simulations with high $p$\ce{CO2}, cooling by Rayleigh scattering begins to outweigh greenhouse warming, such that temperature eventually begins to decrease while \ce{CO2} increases (Fig.~\ref{fig:Seff_OLR_albedo}): in panel A, starting from the lowest $p$CO$_2 = 1$ bar, each contour of $S_{\rm eff}$ moves to higher temperatures as \ce{CO2} increases, until a peak $T_{\rm surf}$ is reached at a threshold $p$\ce{CO2}, beyond which $T_{\rm surf}$ for a given $S_{\rm eff}$ begins to decrease as \ce{CO2} increases. For instance, for $S_{\rm eff} = 0.4$ this peak is at $\approx$280 K and $p$CO$_2 = 10$ bar. This occurs because, at high $p$\ce{CO2}, the albedo (panel C) increases more rapidly with $p$\ce{CO2} than the OLR (panel B) decreases, since the atmosphere has become optically thick at almost all IR wavelengths. In other words, for any given $S_{\rm eff}$ and background gas composition, there is a maximum temperature that cannot be exceeded by adding \ce{CO2} to the atmosphere. This effect has been used to define the outer edge of the classical liquid water habitable zone as the lowest instellation at which atmospheres with 1 bar N$_2$, saturated H$_2$O, and variable \ce{CO2} can maintain an Earth-like planet's global-mean surface temperature above freezing \citep{kasting93,Kopparapu:2013}, i.e. the $S_{\rm eff}$ where the peak temperature is $T_{\rm surf}=273.15$ K. Using this ``maximum greenhouse limit,'' the outer edge of the liquid water habitable zone has been placed at 1.67 astronomical units (au) \citep{Kopparapu:2013} , implying $S_{\rm eff} = \frac{1}{1.67^2} = 0.359$, with $p$\ce{CO2}$\approx$6-7 bar. Our fiducial simulations lack N$_2$, but, in comparison with \citet{Kopparapu:2013}, produce a similar value of $S_{\rm eff} = 0.373$ for the lowest instellation where $T_{\rm surf}$ can be maintained above freezing, with $p$\ce{CO2}$\approx$5-6 bars, demonstrating our climate model produces comparable results to previous efforts.

At low values of $S_{\rm eff}$ (e.g. along the white-dotted $S_{\rm eff}=0.40$ contour in Fig.~\ref{fig:Seff_OLR_albedo}A), we find that the climate responds smoothly to increases in $p$\ce{CO2}, with $T_{\rm surf}$ first increasing and then decreasing until reaching a $p$\ce{CO2}--$T_{\rm surf}$ combination that allows the \ce{CO2} to condense at the surface. This is the point where a given $S_{\rm eff}$ contour intersects the black line that bounds the bottom-right grey area in Fig.~\ref{fig:Seff_OLR_albedo}A. At this point \ce{CO2} is saturated and our simulations assume that any \ce{CO2} added to the atmosphere simply condenses out onto the surface.

As $S_{\rm eff}$ increases, the maximum temperature climates can reach becomes higher and higher. H$_2$O is saturated in our simulations and, as such, $p$H$_2$O increases exponentially with temperature. Therefore, as $S_{\rm eff}$ increases, water's impact on the climate also becomes more and more prominent. In addition to its well-known greenhouse effect, water can impact planetary albedo via several mechanisms, for instance via cloud and sea ice formation. Less obvious impacts of water on planetary albedo come from its contribution to Rayleigh scattering and its competing contribution to shortwave and near-IR absorption.

At low $p$\ce{CO2} ($\lesssim 3$ bar) and high temperatures where $p$H$_2$O is comparable to $p$\ce{CO2}, our simulations indicate that the Rayleigh scattering effect of water starts to become important, which is why the albedo contours bend leftward in the upper-left corner of Fig.~\ref{fig:Seff_OLR_albedo}C, indicating an increase in albedo from enhanced H$_2$O Rayleigh scattering as temperature increases. However, at high enough \ce{CO2} ($\gtrsim$ 1.5 bar), $p$\ce{CO2} remains much larger than $p$H$_2$O even at the highest temperatures we simulated, and hence the Rayleigh scattering effect of H$_2$O is outweighed by that of \ce{CO2}. This suggests that increases to surface temperature up to 360 K stop significantly increasing a planet's Rayleigh scattering albedo via H$_2$O accumulation at $p$\ce{CO2} above a few bar.

Although H$_2$O's Rayleigh scattering ceases to be important at high $p$\ce{CO2}, H$_2$O's shortwave and near-IR absorption remain important, such that the elevated water content caused by increased temperature leads to increased absorption of instellation: albedo and equilibrium $S_{\rm eff}$ decrease with increased temperature at high $p$\ce{CO2}, as indicated by the rightward tilt of $S_{\rm eff}$ and albedo contours in the upper-right quadrants of Fig.~\ref{fig:Seff_OLR_albedo} A and C. In other words, at high $p$\ce{CO2}, H$_2$O saturation leads to temperature-dependent planetary albedo similar to that caused by the ice-albedo feedback, though occurring at temperatures higher than those where the ice-albedo feedback is relevant. This temperature-dependent instellation absorption by H$_2$O introduces a form of hysteresis into the climate system that is analogous to the hysteresis caused by the ice-albedo feedback \citep{abbot2018decrease}, the consequences of which we explore here. 

\subsection{\ce{CO2} ocean oscillations on temperate exoplanets}
\label{subsec:limit_cycling}
Climate limit cycling usually refers to the potential for climates to oscillate back and forth between snowball and temperate states \citep{Menou2015,haqq2016limit,abbot2016analytical,paradise2017}. That occurs when a temperate planet on which weathering dominates outgassing has its \ce{CO2} drawn down until the ice-albedo feedback triggers global glaciation. At this point weathering slows below the rate of outgassing and allows \ce{CO2} to accumulate and eventual deglaciate the planet, which restarts the cycle. Our simulations indicate the existence of a distinct limit cycle that can emerge when outgassing dominates weathering: the oscillation between a \ce{CO2} surface-condensing state and a non-\ce{CO2}-condensing state on planets with \ce{CO2}-H$_2$O atmospheres. This variety of limit cycling is a consequence of the temperature-dependent planetary albedo that arises from H$_2$O's shortwave and near-IR absorption, in combination with \ce{CO2}'s Rayleigh scattering effect.

\begin{figure*}[htb]
    \centering
    \makebox[\textwidth][c]{\includegraphics[width=400pt,keepaspectratio]{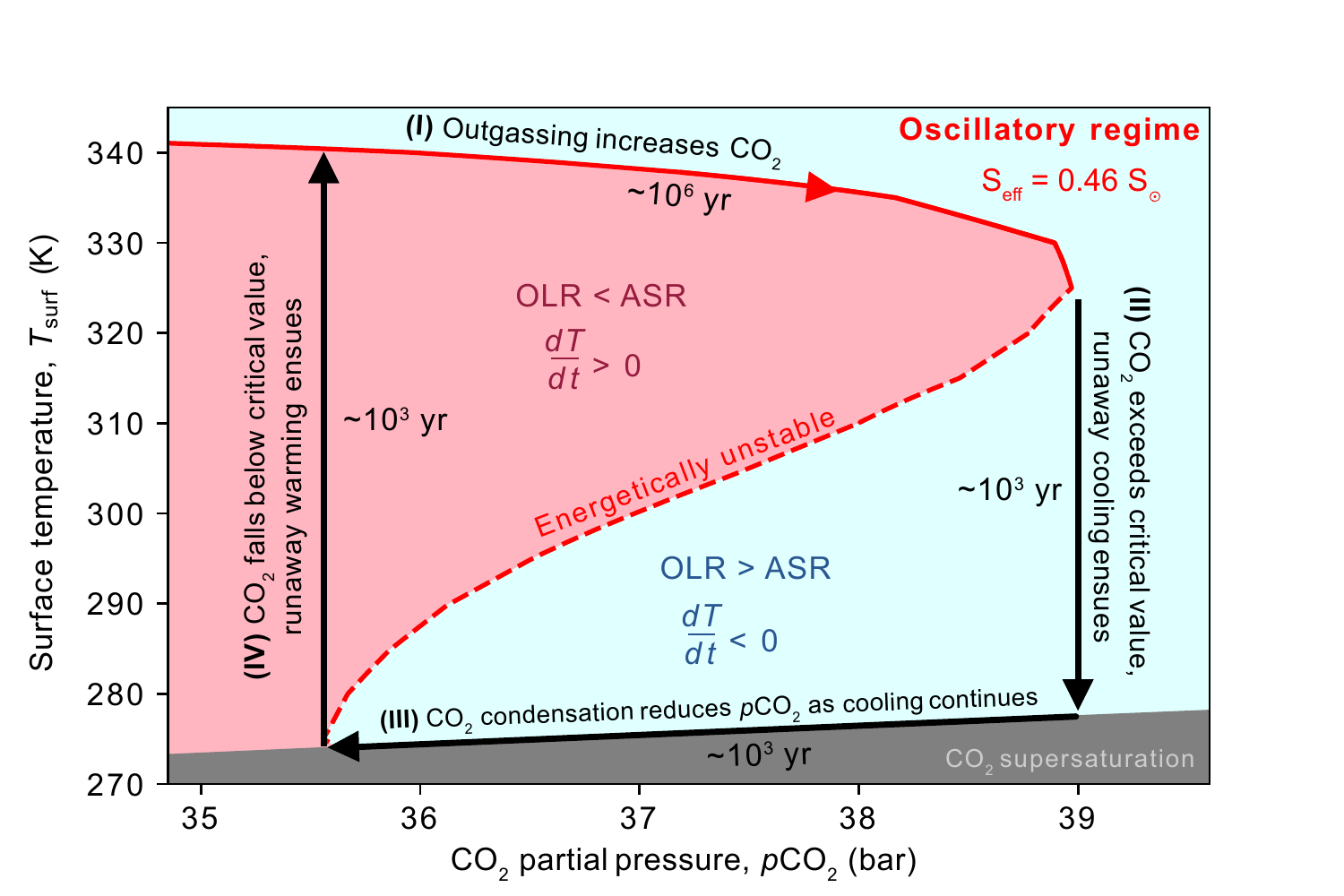}}
    \caption{Climatic limit cycling driven by \ce{CO2} outgassing, Rayleigh scattering, temperature-dependent near-IR stellar absorption by H$_2$O, and surface \ce{CO2} condensation. In step (I), outgassing allows \ce{CO2} to accumulate. In step (II), $p$\ce{CO2} exceeds the value that triggers a positive feedback loop between cooling surface temperature and increasing planetary albedo. In step (III), the planetary surface temperature becomes low enough to trigger surface \ce{CO2} condensation, so continued cooling now also decreases atmospheric $p$\ce{CO2}. In step (IV), a positive feedback loop between warming and albedo reduction from H$_2$O accumulation leads to runaway warming.}
    \label{fig:limitcycles}
\end{figure*}

In equation \ref{eqn:energetic_equilibrium}, energetic equilibrium between global-mean absorbed instellation and OLR is assumed, and each $S_{\rm eff}$ contour in Fig.~\ref{fig:Seff_OLR_albedo}A is a set of $T_{\rm surf}$-$p$\ce{CO2} pairs where equation \ref{eqn:energetic_equilibrium} holds for that particular $S_{\rm eff}$ value. However, at high $p$\ce{CO2}, when $S_{\rm eff}$ is large enough to permit the high temperatures that raise water's vapor pressure enough to lower the planet's albedo substantially, the right-hand branches of the $S_{\rm eff}$ contours become energetically unstable to perturbations in temperature and $p$\ce{CO2}. The consequences of this phenomenon for planetary climate evolution are illustrated in Fig.~\ref{fig:limitcycles}. In this figure, we plot the right-hand branch of the set of $T_{\rm surf}$ and $p$\ce{CO2} values that produce energetic equilibrium with $S_{\rm eff}=0.46$. Any combination of $T_{\rm surf}$ and $p$\ce{CO2} not falling on the dark red line in Fig.~\ref{fig:limitcycles} leads to energetic disequilibrium under an instellation of $S_{\rm eff}=0.46$. This results in either cooling in the case where the planetary outgoing longwave radiation (OLR) is higher than the absorbed stellar radiation (OLR $>$ ASR $=(1-\alpha)S/4$) in the light blue region, or warming in the light red region, where ASR $>$ OLR. 

When the climate resides on the upper, solid portion of the red curve, the temperature responds to energetic disequilibrium as follows: at a given $p$\ce{CO2}, a perturbation in temperature upward from equilibrium is met with a cooling response and a perturbation in temperature downward is met with a warming response, until energetic equilibrium is re-established. Similarly, an increase in \ce{CO2} from equilibrium increases albedo, which causes the planet to cool to maintain equilibrium. Thus small changes to $T_{\rm surf}$ or $p$\ce{CO2} in this region of parameter space near the solid red curve are met with a negative feedback that tends to maintain energetic equilibrium and move the climate back to the red curve. Starting from somewhere on the red curve near the area labeled (I), if \ce{CO2} in the atmosphere is increased by outgassing, moving rightward along the equilibrium curve, the climate will eventually reach the curve's rightmost point.

If $p$\ce{CO2} is increased beyond the value at the rightmost point of the red curve, the increase in albedo is enough to push the climate system into a state of energetic disequilibrium where OLR $>$ ASR (the light blue region in Fig.~\ref{fig:limitcycles}), and the planet begins to cool. In this region, OLR responds only weakly to changes in temperature because $p$\ce{CO2} is so high that the atmosphere is mostly opaque in the IR (see Fig.~\ref{fig:Seff_OLR_albedo}B), but the albedo responds substantially, increasing as temperature decreases, since H$_2$O in the atmosphere falls exponentially with temperature, reducing the atmosphere's ability to absorb instellation. This produces a positive feedback with runaway cooling (stage (II) in Fig.~\ref{fig:limitcycles}), where a reduction in temperature dries the atmosphere, which decreases ASR and thus pushes the system even further out of energetic equilibrium. This accelerates the cooling and reduces the ASR further. Any plausible rate of cooling vastly exceeds plausible rates of \ce{CO2} accumulation from outgassing (compare thermal equilibration timescale of 1000 years \citep{Pierrehumbert:2010-book} to a carbon cycle timescale of $10^6$ years or more \citep{colbourn2015time}), so the cooling trajectory is effectively straight down in $T_{\rm surf}$-$p$\ce{CO2} space.

After a temperature reduction of approximately 50 K, the conditions at the surface have cooled enough for \ce{CO2} to begin to condense out onto the planetary surface, which is in this case covered by a liquid H$_2$O ocean. In stage (III) of the climate cycle, $p$\ce{CO2} is directly dictated by surface temperature via \ce{CO2}'s Clausius Clapeyron relation. Since the climate is still in the light blue region where OLR$>$ASR, continued cooling drives a rapid decrease in \ce{CO2} partial pressure, resulting in decreasing planetary albedo and increasing ASR.

Eventually, the reduction in \ce{CO2} partial pressure increases ASR enough to re-equilibrate with the OLR at the point where arrow (III) meets arrow (IV) in Fig.~\ref{fig:limitcycles}. However, this equilibrium point is unstable to further reductions in \ce{CO2} or increases to temperature, which would shift the climate into the light red region of Fig.~\ref{fig:limitcycles}, where OLR$<$ASR and warming is self-reinforcing due to the accumulation of atmospheric water vapor and resultant reduction in planetary albedo. As a result, any internal climate variability that acted to transiently warm the climate away from this unstable equilibrium would trigger the positive warming feedback loop represented by arrow (IV) in Fig.~\ref{fig:limitcycles}, analogous to the cooling feedback loop represented by arrow (II).  This warming feedback loop would finally carry the climate back to its initial energetically-stable state, at about 35.6 bar of \ce{CO2} and $T_{\rm surf} \approx$ 340 K. From here, assuming outgassing continues, the planet would begin another iteration of this cycle of atmospheric \ce{CO2} accumulation, runaway cooling, \ce{CO2} rain-out, and runaway warming. 

For this example of a \ce{CO2} ocean cycle, we chose an instellation that kept the surface temperature at each $p$\ce{CO2} within the range of temperatures we simulated ($\leq$ 365 K). With higher $S_{\rm eff}$, the maximum attainable temperature increases, and the unstable righthand branches of the $S_{\rm eff}$ contours shift rightward to higher $p$\ce{CO2} (see Fig.~\ref{fig:Seff_OLR_albedo}). Both of those responses to higher $S_{\rm eff}$ would increase the size of temperature jumps over the course of a limit cycle. Therefore, planets that start off in a limit cycling state at low $S_{\rm eff}$ will undergo cycles of greater and greater amplitude as their star brightens and incident instellation increases.

Up to this point, we have discussed the evolution of climates with \ce{CO2} outgassing but without a complementary weathering feedback. This can correspond to a scenario in which weathering is ``supply-limited,'' i.e. the supply of weatherable minerals to the planetary surface is too low for weathering to keep up with the rate of \ce{CO2} outgassing, or a scenario where liquid \ce{CO2} or \ce{CO2} clathrate hydrate blankets the ocean floor, suppressing weathering reactions in seafloor basalts (a scenario discussed further in Section \ref{sec:discussion}). In the next subsection, we present calculations that include a simple weathering feedback.

\begin{figure*}[htb]
    \centering
    \makebox[\textwidth][c]{\includegraphics[width=400pt,keepaspectratio]{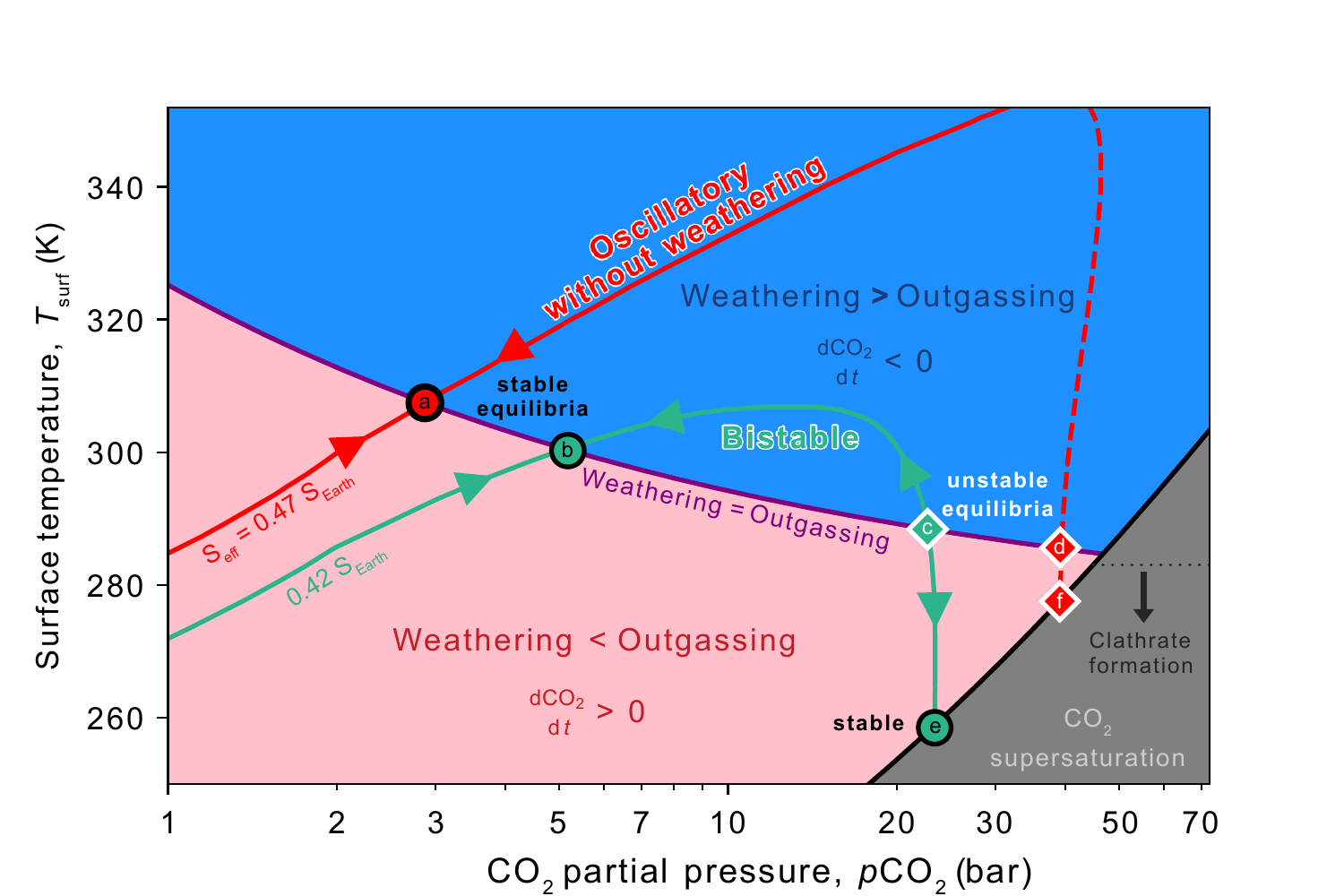}}
    \caption{Comparison of stable (circles with black outline: a, b, e) and unstable (diamonds with white outline: c, d, f) climate equilibria at different instellations. The red line represents climate states with $S_{\rm eff}=0.47 \; S_{\rm Earth}$, with the solid part of the red line representing energetically stable climates and the dashed part of the red line representing energetically unstable climates. The green line represents climates with $S_{\rm eff}=0.42 \; S_{\rm Earth}$, with the possibility of bistability between an Earth-like scenario and a CO$_2$ condensing state. The purple line shows where weathering is equal to outgassing; the light red region below is where outgassing outpaces weathering, and the light blue region above is where weathering outpaces outgassing.}
    \label{fig:equilibria}
\end{figure*}

\subsection{Bistability from the interaction of Rayleigh scattering, weathering, and \ce{CO2} condensation}
\label{subsec:bistability}

With weathering included (Eq. \ref{eqn:mac_weathering}) and assuming an outgassing rate of 15.8$\times10^{12}$ mol yr$^{-1}$ (2.1$\times$ an Earth-like rate of 7.5$\times 10^{12}$ mol yr$^{-1}$ drawn from \citet{haqq2016limit}), our model produces a set of weathering-outgassing equilibria corresponding to the solid purple line in Fig.~\ref{fig:equilibria}. The curve has a negative slope because of the $p$\ce{CO2}-dependence of weathering, with lower $T_{\rm surf}$ required for weathering/outgassing equilibrium at higher values of $p$\ce{CO2}. A larger outgassing rate would result in the purple line residing at higher temperatures for a given $p$\ce{CO2}, changing the locations of stable and unstable equilibria, and a smaller outgassing rate would have the opposite effect. With a large enough increase in outgassing (just an increase to 2.2$\times$ the Earth-like rate, in this case), the low-CO$_2$ solution becomes inaccessible, and with a large enough decrease in outgassing (a reduction to below 1.0$\times$ the Earth-like rate, in this case), the high-CO$_2$ solution similarly disappears. Changes to the parameters in the weathering model (for example, changing the assumed global-mean soil thickness) would have analogous effects on the locations and presence of the equilibria. With the formulation of weathering applied here \citep[from][]{maher2014hydrologic,winnick2018relationships,graham2020thermodynamic}, the behavior of the system is quite sensitive to changes in outgassing, land fraction, or weathering parameters, while the more traditional kinetically-limited formulation introduced in \citet{Walker-Hays-Kasting-1981:negative} would result in less sensitivity, as demonstrated in \citet{graham2020thermodynamic}.

Fig.~\ref{fig:equilibria} indicates the existence of two equilibrium points where outgassing ($V$) is balanced by weathering ($W$), $V=W$, and OLR = ASR for planets at a given $S_{\rm eff}$, with one equilibrium climate having a higher $T_{\rm surf}$ and a lower $p$\ce{CO2} (red and green circles $a$ and $b$ in Fig.~\ref{fig:equilibria}) than the other (red and green diamonds $c$ and $d$). The physical reason for pairs of equilibria at each $S_{\rm eff}$ is \ce{CO2}'s cooling effect at high partial pressures. For both $S_{\rm eff}$ values (0.42 and 0.47 $S_{\rm Earth}$, green line and red line) plotted in Fig.~\ref{fig:equilibria}, the warmer, lower-$p$\ce{CO2} climate equilibrium ($a$ and $b$) is stable with respect to both its energy fluxes and its carbon fluxes, meaning a planet will return to that climate equilibrium if perturbed away from it. These stable climates are the equilibria that are typically explored in studies of silicate weathering on terrestrial planets. 

Unlike its counterpart, the second equilibrium climate state where $V=W$ on each $S_{\rm eff}$ curve ($c$ and $d$) is unstable to climatic perturbations. This is best illustrated by examining unstable equilibrium $c$ with $S_{\rm eff}=0.42 \; S_{\rm Earth}$ in Fig.~\ref{fig:equilibria}. As noted earlier, compared to the timescale of carbon cycle response, the thermal equilibration timescale is instantaneous, so for this discussion we can assume that the climate is constrained to move along the green curve at all times. If a planet begins with a climate at point $c$, and its $p$\ce{CO2} is perturbed downward (leftward on the plot), it warms up because of a reduction in \ce{CO2} Rayleigh scattering and moves upward along the green $S_{\rm eff}$ isoline. This moves the planet into the blue zone of Fig.~\ref{fig:equilibria} where $W>V$, which means that \ce{CO2} is now being consumed by weathering faster than it can be supplied by outgassing. This imbalance in carbon fluxes leads to further reduction in $p$\ce{CO2}, enhancing the initial perturbation and pushing the climate deeper and deeper into the blue region. Eventually the planet reaches the peak temperature for that $S_{\rm eff}$, at which point the continued reduction in \ce{CO2} begins to cool the climate, slowing weathering until finally the planet reaches the stable equilibrium point $b$.

Conversely, if a planet begins on the unstable equilibrium $c$ and $p$\ce{CO2} is perturbed upward (to the right on the plot), the planet's surface will cool and the climate will move into the light red area of Fig.~\ref{fig:equilibria}, where outgassing is greater than weathering ($V>W$). With outgassing now outpacing weathering, $p$\ce{CO2} will continue to grow, enhancing the initial climate perturbation until the $T_{\rm surf}$--$p$\ce{CO2} combination allows for \ce{CO2} condensation at the planetary surface, at which point the carbon cycle has reached a new, stable equilibrium $e$ governed by equation \ref{eqn:cond_flux}, with the imbalance between outgassing and weathering being balanced by surface condensation of \ce{CO2}. This suggests that rocky planets at low instellation can display carbon cycle bistability, where the same geologic boundary conditions (as represented by the parameters in equation \ref{eqn:mac_weathering}) and same stellar environment can drive two very different stable equilibrium climates with $p$\ce{CO2} differing by an order of magnitude, one of which displays surface \ce{CO2} condensation and one of which does not.

The carbon cycling behavior of a planet irradiated by relative instellation of $S_{\rm eff}=0.47 \; S_{\rm Earth}$ (the green curve in Fig.~\ref{fig:equilibria}) is considerably different than that of the previous example because of the H$_2$O absorption-based radiative feedbacks that cause the CO$_2$ ocean limit cycling discussed in Section \ref{subsec:limit_cycling}. In this case, even though the intersection point $f$ (red-white diamond) between the $S_{\rm eff}=0.47 \; S_{\rm Earth}$ curve (red, dashed curve) and the \ce{CO2} saturation vapor pressure curve (black) is stable with respect to the carbon cycle, it is unstable with respect to energy fluxes: a small perturbation in temperature upward or \ce{CO2} downward from that point would trigger self-reinforcing warming like that exemplified by arrow (IV) in Fig.~\ref{fig:limitcycles}. This would warm the planet to $T_{\rm surf}\approx350$ K. This is in the blue region of Fig.~\ref{fig:equilibria}, where weathering outpaces outgassing, so \ce{CO2} would subsequently be consumed by weathering until the carbon cycle reached the stable equilibrium $a$, where $V=W$ for $S_{\rm eff} = 0.47 \; S_{\rm Earth}$. 

In summary, within the modeling framework applied in this article, a climate transition occurs between $S_{\rm eff}=0.42 \; S_{\rm Earth}$ and $S_{\rm eff}=0.47 \; S_{\rm Earth}$ where the climate configurations that allow for surface \ce{CO2} condensation become energetically unstable by the same mechanism that allows for the CO$_2$ ocean limit cycles described in Section \ref{subsec:limit_cycling}. However, in this case, the addition of a weathering feedback terminates the cycle before its \ce{CO2} accumulation phase (analogous to step (I) in Fig.~\ref{fig:limitcycles}) can be initiated.

\subsection{Variations in stellar type}
As noted in Section \ref{subsec:rad}, the fiducial case studied in this article is irradiated by a solar (G2V type) spectrum drawn from \citet{Kurucz1995}. Irradiation by spectra corresponding to different stellar types can result in different climate behavior compared to what we have examined so far. Here we examine the impact of an F-type (F2V) spectrum and an M-type (M3.5V) spectrum on our basic results. In all cases, the OLR remains the same, but the planetary albedo is altered by changes to the stellar spectrum impinging on the planet. The general trend is simple to state: planets orbiting hotter, bluer stars can support the bistability between climates with and without CO$_2$ oceans at higher instellations than cooler, redder stars. 

\subsubsection{F-type stars}

\begin{figure*}[htb]
    \centering
    \makebox[\textwidth][c]{\includegraphics[width=400pt,keepaspectratio]{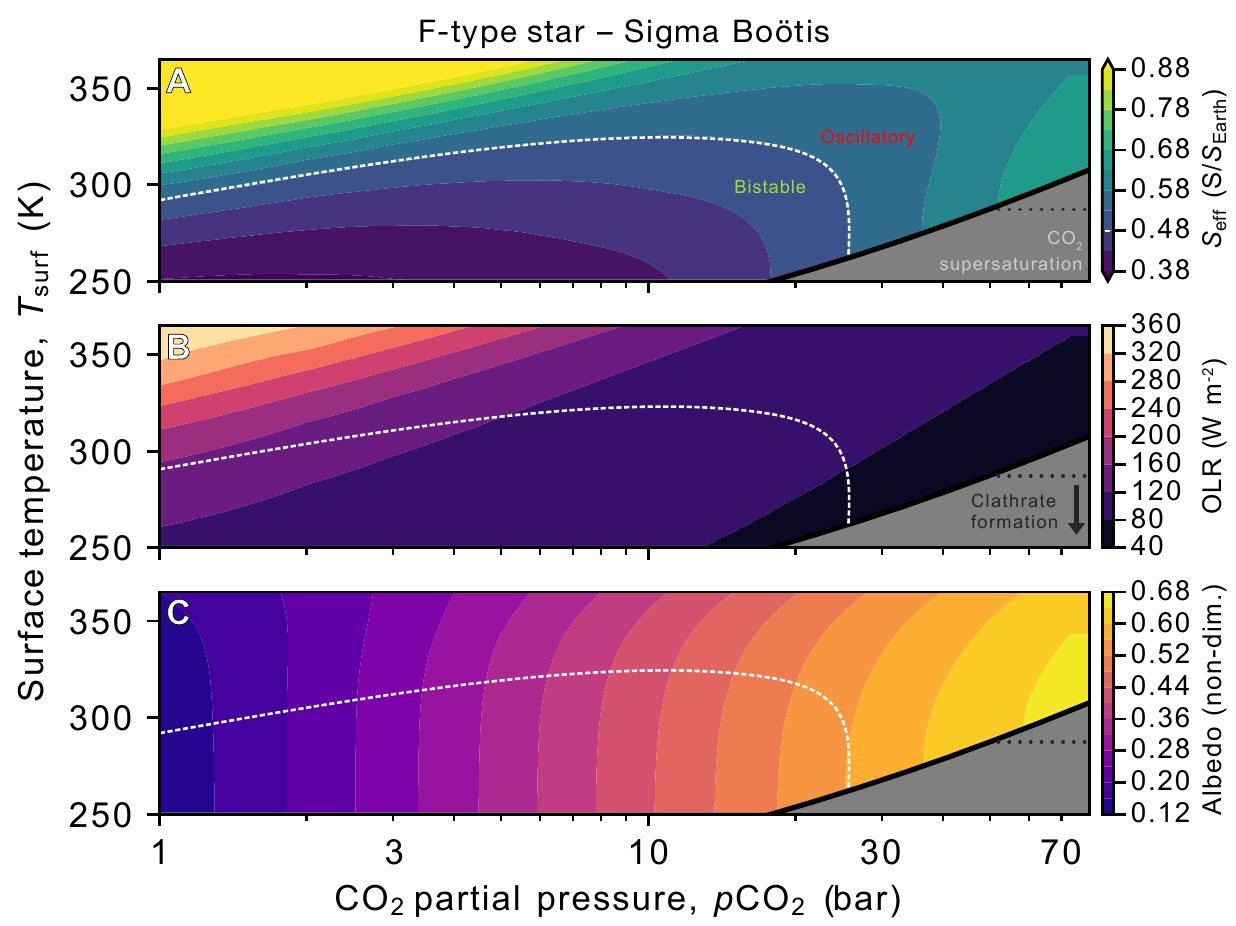}}
    \caption{Energetic properties of terrestrial planetary climates irradiated by Sigma Boötis, an F2V star, at low instellation as a function of surface temperature $T_\mathrm{surf}$ ($y$-axes) and \ce{CO2} partial pressure ($x$-axes). Panels and color coding are the same as in Fig.~\ref{fig:Seff_OLR_albedo}.}
    \label{fig:Fstar}
\end{figure*}

To examine the behavior of climates irradiated by F-type stars, we apply the spectrum of Sigma Boötis, an F2V star, drawn from \citet{segura03}, in Fig.~\ref{fig:Fstar}. Because F-type stars display spectra that are shifted toward higher (bluer) frequencies than G- or M-type stars, the Rayleigh scattering effect of CO$_2$ is stronger for planets orbiting these stars, as Rayleigh scattering increases greatly in efficacy at shorter wavelengths. Thus, a given increase in $p$CO$_2$ leads to a larger increase in albedo for these planets, as demonstrated in Fig.~\ref{fig:Fstar}C, where the planetary albedo reaches nearly 0.68 at the highest $p$CO$_2$ shown (compared to a maximum albedo of 0.56 in simulations irradiated by a solar spectrum shown in Fig.~\ref{fig:Seff_OLR_albedo}C). This higher sensitivity of albedo to $p$CO$_2$ leads to climate configurations where the bistability between climates with and without CO$_2$ oceans can persist to a higher instellation, reaching $\approx54\%$ of that of modern-day Earth (Fig.~\ref{fig:Fstar}A), compared to $\approx45\%$ for climates irradiated by the solar spectrum (Fig.~\ref{fig:Seff_OLR_albedo}). 

\subsubsection{M-type stars}

\begin{figure*}[htb]
    \centering
    \makebox[\textwidth][c]{\includegraphics[width=400pt,keepaspectratio]{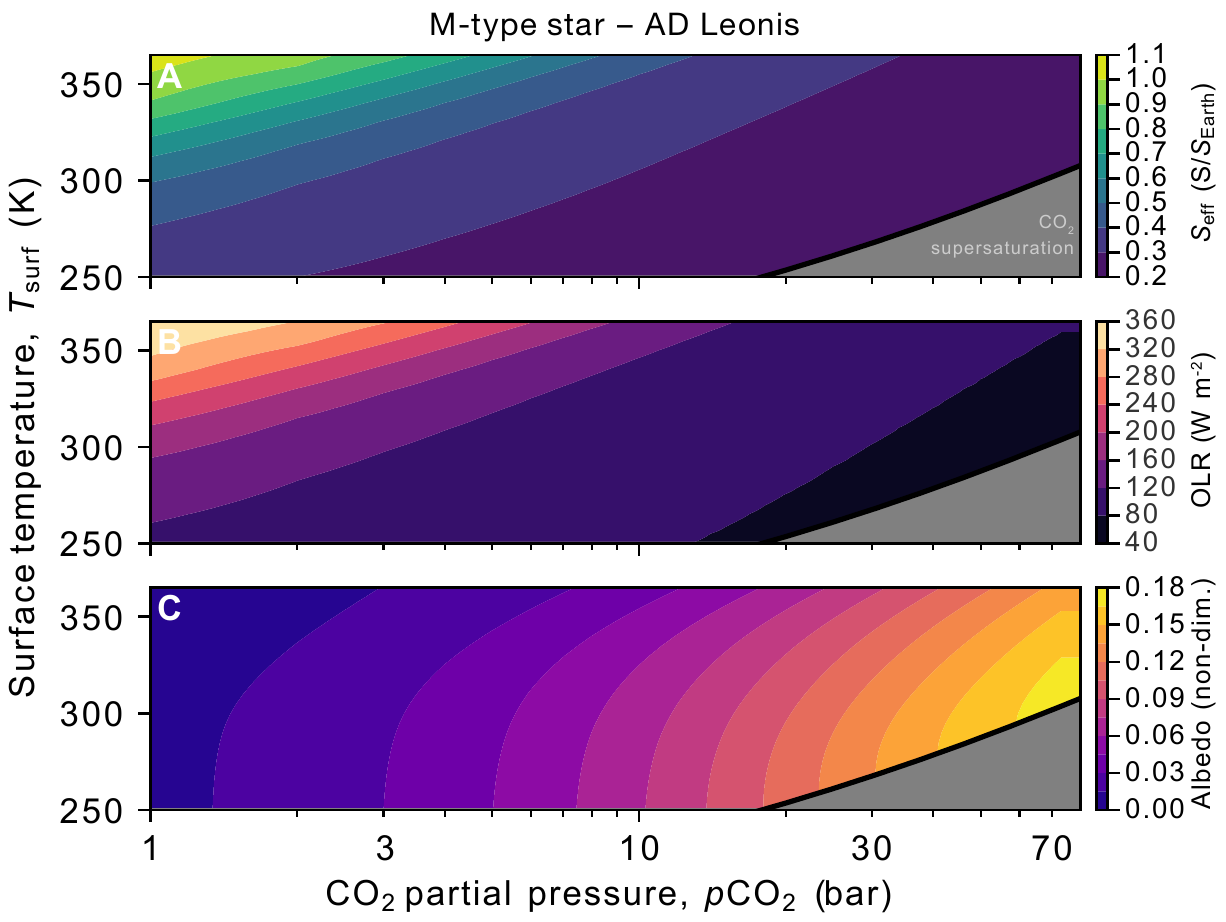}}
    \caption{Energetic properties of terrestrial planetary climates irradiated by AD Leonis, an M3.5V star, at low instellation as a function of surface temperature $T_\mathrm{surf}$ ($y$-axes) and \ce{CO2} partial pressure ($x$-axes). Panels and color coding are the same as in Fig.~\ref{fig:Seff_OLR_albedo}.}
    \label{fig:Mstar}
\end{figure*}

To examine the behavior of planets orbiting M-type stars, we apply the spectrum of AD Leonis, an M3.5V star \citep{Segura:2005}, in Fig.~\ref{fig:Mstar}. For planets orbiting M-type stars, which emit a larger proportion of their energy at lower (redder) frequencies than G- or F-type stars, the Rayleigh scattering impact of atmospheric $p$CO$_2$ is weaker. Even at the highest $p$CO$_2$ simulated in this paper (72 bar), albedo does not exceed 0.18 (Fig.~\ref{fig:Mstar}C) for planets orbiting AD Leonis, compared to maximum simulated albedo values of 0.56 and 0.68 for planets orbiting G- and F-type stars, respectively. 

Interestingly, this weak dependence of planetary albedo upon CO$_2$ leads to a complete lack of bistable climate states in our M-star simulations. Without a substantial increase in albedo with $p$CO$_2$, CO$_2$'s reduction of OLR dominates its radiative impact, allowing the molecule to remain a net greenhouse gas across the entirety of the parameter space we studied. Thus, for planets with functional weathering feedbacks that orbit M-type stars, we expect there to be only one stable climate state, corresponding to an Earth-like equilibrium that lacks condensed CO$_2$ at the surface. Correspondingly, we do not expect planets orbiting M-stars to display limit cycling between CO$_2$-condensing and non-condensing states, regardless of outgassing rate or TOA instellation.

\section{Discussion}\label{sec:discussion}
The interaction of liquid \ce{CO2} and liquid \ce{H2O} at the planetary surface is fundamental to evaluating the climate state and surface geochemistry of planets with large inventories of both condensed phases. The range of temperatures over which liquid \ce{CO2} would be stable in the presence of a liquid water ocean is relatively narrow \citep{marounina2020internal}. At temperatures above 304.5 K \ce{CO2} becomes supercritical and does not condense. At temperatures below 282.91 K and pressures below 4.46 MPa (44.6 bar), mixtures of condensed \ce{CO2}, condensed \ce{H2O}, and \ce{CO2}-rich vapor are metastable, with an equilibrium state of \ce{CO2} hydrate, a crystalline phase where water molecules encase \ce{CO2} molecules \citep{wendland1999experimental}. 

Experiments in Earth's ocean demonstrate that hydrate forms rapidly upon contact between liquid \ce{CO2} and liquid \ce{H2O}, with visible masses forming over the course of just a few hours \citep{brewer1999direct}. This suggests that with surface temperatures below the 282.91 K quadruple point of \ce{CO2} hydrates mentioned in the previous paragraph, precipitation of condensed \ce{CO2} from the atmosphere into a liquid water-rich ocean would result in immediate formation of solid hydrates that would then sink through the water and settle on the seafloor. This may result in large-scale hydrate build-up on the ocean floor, which would suppress or halt seafloor silicate weathering, similar to high-pressure ice phases on waterworlds with extremely deep oceans \citep{kite2018habitability,2020SSRv..216....7J}, removing an important \ce{CO2} sink and making it even more difficult for a planet to exit a stable \ce{CO2} condensing state. Other forms of low-temperature seafloor alteration would also be dramatically altered, with likely major consequences for ocean chemistry \citep{coogan2018low}. Similar layered structures of hydrate and water have also been proposed for icy moons and dwarf planets \citep{bostrom2021self}, suggesting exploration of such bodies in the solar system may also provide insight into the structure of low-instellation terrestrial planets. Further, assuming a slow rate of subduction, the formation of large seafloor hydrate reservoirs could consume large fractions of the water in planets with Earth-like volatile inventories, since each \ce{CO2} molecule in a hydrate is accompanied by 5.75 \ce{H2O} molecules in the most common hydrate structure \citep{brewer1999direct}. Thus, under the conditions where \ce{CO2} hydrates are stable (at temperatures below 282.91 K), a planet's subduction rate could exert a powerful direct control on ocean depth and salinity, both of which are first-order parameters in determining planetary climate and surface geochemistry \citep{olson2020oceanographic}. Finally, the coexistence of liquid \ce{CO2} and \ce{H2O} in the air may lead to the formation of aerial \ce{CO2} hydrates, altering the atmospheric lapse rate through latent heat release \citep{kasting1991co2}, but it is unclear whether this would be an efficient process given the factor-of-a-thousand difference in vapor pressure between \ce{CO2} and \ce{H2O} at relevant temperatures.

At temperatures above the hydrate quadruple point (282.9 K) and below \ce{CO2}'s triple point (304.5 K), liquid \ce{CO2} and liquid \ce{H2O} can coexist. Under pressures like those at the sea surface, liquid \ce{CO2} is less dense than liquid H$_2$O, so \ce{CO2} that rains into the ocean from the atmosphere would float and form a layer on top of the water \citep{house2006permanent,marounina2020internal}. However, liquid \ce{CO2} is also more compressible than liquid water, such that \ce{CO2} can become denser than water at high enough pressures \citep{house2006permanent,marounina2020internal}. This means that if liquid \ce{CO2} enters the water ocean at a deep enough point, then instead of floating on top of the water, it will sink to the seafloor. Therefore, submarine \ce{CO2} degassing in such conditions would result in the formation of a liquid \ce{CO2} layer at the lithosphere/ocean interface. It is unclear whether silicate weathering by liquid \ce{CO2} is possible, so the impact of this seafloor layer on carbon sequestration is an open question. All in all, the coexistence of liquid \ce{CO2} and liquid \ce{H2O} suggests the possibility of ``layer cake'' oceans at the surfaces of some rocky, low-instellation planets, with an \ce{H2O} layer nestled between two \ce{CO2} layers. This analysis remains speculative without detailed thermodynamic modeling, which is beyond the scope of this article but would make for insightful work on the surface conditions of rocky exoplanets under low instellation.

The properties of the atmosphere and ocean under \ce{CO2} condensing conditions also have implications for the viability of origin of life scenarios on prebiotic planets akin to the Hadean Earth. In one popular school of thought regarding the origin of life, surface UV fluxes are held to be important drivers of prebiotic chemistry \citep{sasselov2020origin,2021NatCh..13.1126L}. The surface UV flux has been quantified in a variety of models approximating early Earth and early Mars atmospheres  \citep{rugheimer2015uv,ranjan2016influence,ranjan2017atmospheric}, and atmospheres with multibar CO$_2$ pressures demonstrated significant UV attenuation from scattering and absorption \citep{ranjan2017atmospheric}. This suggests that atmospheres with 10s of bar of CO$_2$ like those expected for the CO$_2$ condensing cases examined in this paper would not receive enough UV light at their surfaces to drive the relevant prebiotic chemistry. With respect to submarine origin of life scenarios, the seafloor being covered in liquid \ce{CO2} or \ce{CO2} hydrate in conditions with surface \ce{CO2} condensation would preclude water-rock reactions like serpentinization \citep{sleep2011serpentinite} or aqueous organo-metal chemistry near hot vents \citep{sobotta2020possible}. It seems that several major models for the origin of life require processes that would be difficult on planets displaying the \ce{CO2} condensing conditions explored here, even with temperate and otherwise habitable surface climates. 

Given the likely reduced potential for life to emerge on CO$_2$ condensing planets, an observational discriminant between these worlds and their non-condensing counterparts would be useful for prioritizing targets in the search for life beyond the solar system. Here, we propose a potential method for distinguishing between CO$_2$ condensing and non-condensing worlds and, more generally, constraining the $p$\ce{CO2} of a given exoplanet. This proposal is based on the tendency of \ce{CO2} to dimerize and form molecular complexes at high pressures and low temperatures like those that occur in the \ce{CO2} condensing atmospheres explored in this paper \citep{leckenby1966observation,slanina1992computational,tsintsarska2007equilibrium,asfin2015communication}. Similar to a method proposed for constraining \ce{O2} partial pressure using spectroscopic \ce{O2} dimer features in exoplanet observations \citep{misra2014using}, we suggest that high resolution spectroscopy from astronomical surveys may be able to detect \ce{CO2} dimer features on high-$p$\ce{CO2} exoplanets. Using a quadratic fit to estimate the CO$_2$ dimerization equilibrium constant as a function of temperature based on data from molecular dynamics simulations \citep{tsintsarska2007equilibrium}, we find that \ce{CO2} dimers would make up $\approx$5\% of the atmosphere by molar fraction at the surface on planets with surface CO$_2$ condensation and surface temperatures between 273.15 K (\ce{H2O}'s freezing point temperature) and 304 K (\ce{CO2}'s critical temperature). Planets in the same temperature range but with lower, non-condensing $p$\ce{CO2} levels have much smaller levels of dimerization; for example, an atmosphere with 6 bar of CO$_2$ and a surface temperature of 288 K would have a dimer fraction of less than a percent at the surface. Consequently, it may be possible to discriminate between CO$_2$-condensing and non-CO$_2$-condensing atmospheres using the presence or absence of CO$_2$ dimer features. These features have been detected in the near- and mid-infrared around 3700 cm$^{-1}$ (2.7 $\mu$m) \citep{jucks1987structure,jucks1988structure,moazzen2013spectroscopy}, 2350 cm$^{-1}$ (4.3 $\mu$m) \citep{walsh1987pulsed,dehghany2010high}, and 1250-1400 cm$^{-1}$  (7.1--8.0 $\mu$m) \citep{baranov2004infrared, asfin2015communication}, placing the features within the proposed wavelength range of future telescope architectures like the Large Interferometer For Exoplanets \citep{quanz2021atmospheric,2022arXiv220300471D}. \citet{fox1988spectra} note the possibility of detection of CO$_2$ dimers in the atmospheres of Mars and Venus, where they are suggested to exist at parts-per-thousand mole fractions, much lower than the expected dimer abundance in CO$_2$ condensing atmospheres due to the low pressure and high temperatures on Mars and Venus respectively. 

If the proposed method for constraining $p$\ce{CO2} by dimer feature detection is borne out under more comprehensive examination, it may allow identification of statistical trends in $p$\ce{CO2} versus instellation. Such a trend is expected if there is a population of Earth-like planets in the habitable zones of stars with \ce{CO2} levels controlled by silicate weathering, which would introduce a trend of \ce{CO2} decrease with increasing instellation \citep{bean2017statistical,checlair2019statistical,lehmer2020carbonate}. Conversely, for the population of planets where CO$_2$ levels are controlled by condensation, the opposite trend would emerge: $p$\ce{CO2} would increase with increasing instellation. For instance, this trend can be observed in Fig.~\ref{fig:Seff_OLR_albedo}A, where \ce{CO2}'s saturation vapor pressure curve intersects increasingly large $S_{\rm eff}$ contours as $p$\ce{CO2} increases. Statistical comparative planetology may thus be able to distinguish populations of terrestrial exoplanets with \ce{CO2} levels controlled by different physical and chemical processes, even if the individual measurements are too low in precision to unambiguously place a given planet into either population. A more detailed analysis of spectral response for the different climate scenarios we outline will be beneficial to analyse the optimal observational architecture \citep{LUVOIR_StudyReport2019,2019AJ....158...83A,LIFE2021a}.

\subsection{Caveats}
Any proposal to analyze the atmospheres of exoplanets is hindered by the potential presence of clouds, and the above is no different: high altitude cloud decks consisting of liquid or solid CO$_2$ or H$_2$O may obscure parts of the atmosphere on these \ce{CO2} condensing planets. Clouds might also impact the OLR and albedo of these planets, potentially changing the patterns of climate behavior as a function of $p$\ce{CO2} and instellation. Thus, the inclusion of clouds may alter the conclusions of this clear-sky study. In particular, if it turns out that thick, global, high-altitude cloud decks obscure the bulk of the planetary atmosphere in most high-$p$CO$_2$ climates, the bistability we identify might be muted or eliminated due to the reduced importance of CO$_2$ Rayleigh scattering of visible light under such conditions. Planetary albedo would instead be determined by cloud properties, with little dependence on $p$CO$_2$. We do not include clouds in our simulations for a variety of reasons. Most importantly, it is simply not possible to self-consistently calculate realistic cloud distributions in a one-dimensional model, or even to theoretically estimate cloud deck locations or cloud condensation nuclei density, so any attempt at cloud inclusion would either require arbitrary choices of all of these fundamental parameters for four distinct cloud varieties (solid H$_2$O, liquid H$_2$O, solid CO$_2$, liquid CO$_2$) or a many-dimensional parameter space sweep. Further, to our knowledge, the physical and optical properties of liquid \ce{CO2} droplets have not been measured, making it difficult to estimate their impact without making unsupported guesses about their physical properties. The radiative impacts of \ce{CO2} ice clouds have been examined in some detail \citep{Forget:1997p3442,forget20133d,kitzmann2016revisiting, kitzmann2017clouds}, and most recent work has found that their net effect on climate is likely to be small under most parameter assumptions, though further work is warranted on this problem. The behavior of water clouds (either solid or liquid) in thick CO$_2$ atmospheres is relatively under-explored, though some 3D GCM studies of early Mars \citep{Wordsworth:2013fk,kite2021warm} have examined this regime. \citet{Wordsworth:2013fk} found a small radiative impact from water clouds in the cool, arid climates they simulated. \citet{kite2021warm} found a significant climate effect from high-altitude water clouds in arid simulations and a minimal effect in simulations with a global ocean. Our simulations assumed an Earth-like global ocean. Thus, although clouds are a significant source of uncertainty in climate modeling, excluding their effects for a principal examination \citep[similar to studies on the runaway greenhouse effect, e.g.][]{Kopparapu:2013} of the phenomena we are studying is justified. 

We also neglected the ice-albedo feedback in our simulations because the vast majority of climates we examined had surface temperatures above freezing. Previous studies \citep[e.g.][]{turbet2017co,kadoya2019outer} have found that planets that fall into globally glaciated states at low instellation may experience surface CO$_2$ condensation at drastically lower $p$CO$_2$ values than the planets we have examined, since the surface temperatures of glaciated planets are tens of Kelvin colder than surface temperatures on temperate planets, though a functional seafloor weathering feedback might be enough to draw down CO$_2$ to low levels and prevent CO$_2$ condensation even in a snowball state at low instellations \citep[e.g.][]{chambers2020effect}. The potential for a climate transition from a snowball state with surface CO$_2$ condensation to a temperate state with continued surface CO$_2$ condensation is an interesting target for further modeling, and may provide an alternative route to a bistable CO$_2$-ocean-bearing climates on Earth-like planets with low instellation. In some regions of parameter space (especially low instellation and high $p$\ce{CO2}), the accumulation of CO$_2$ and resultant cooling from Rayleigh scattering could itself drive a planet into a snowball state as well, providing another intriguing and counter-intuitive climate scenario for follow-up. 

Finally, we note that meridional surface temperature gradients on these planets could lead to CO$_2$ condensation at somewhat lower surface pressures for a given temperature than calculated here, as the poles tend to be cooler than the global mean surface temperature on planets with Earth-like obliquities and rotation rates, allowing CO$_2$ surface condensation with less atmospheric CO$_2$ accumulation. However, meridional surface temperature gradients are greatly reduced at high surface pressures \citep{chemke2017dynamics} and atmospheres that are made up mostly of condensable species also tend to have very small meridional temperature gradients due to the powerful winds that develop in response to the large pressure gradients that would be caused by temperature gradients in condensable-rich atmospheres. For example, the equator-to-pole temperature gradient for a pure H$_2$O atmosphere is calculated to be on the order of $\sim1$ K \citep{ding2018global}.

\begin{figure*}
    \centering
    \makebox[\textwidth][c]{\includegraphics[width=1.29\textwidth]{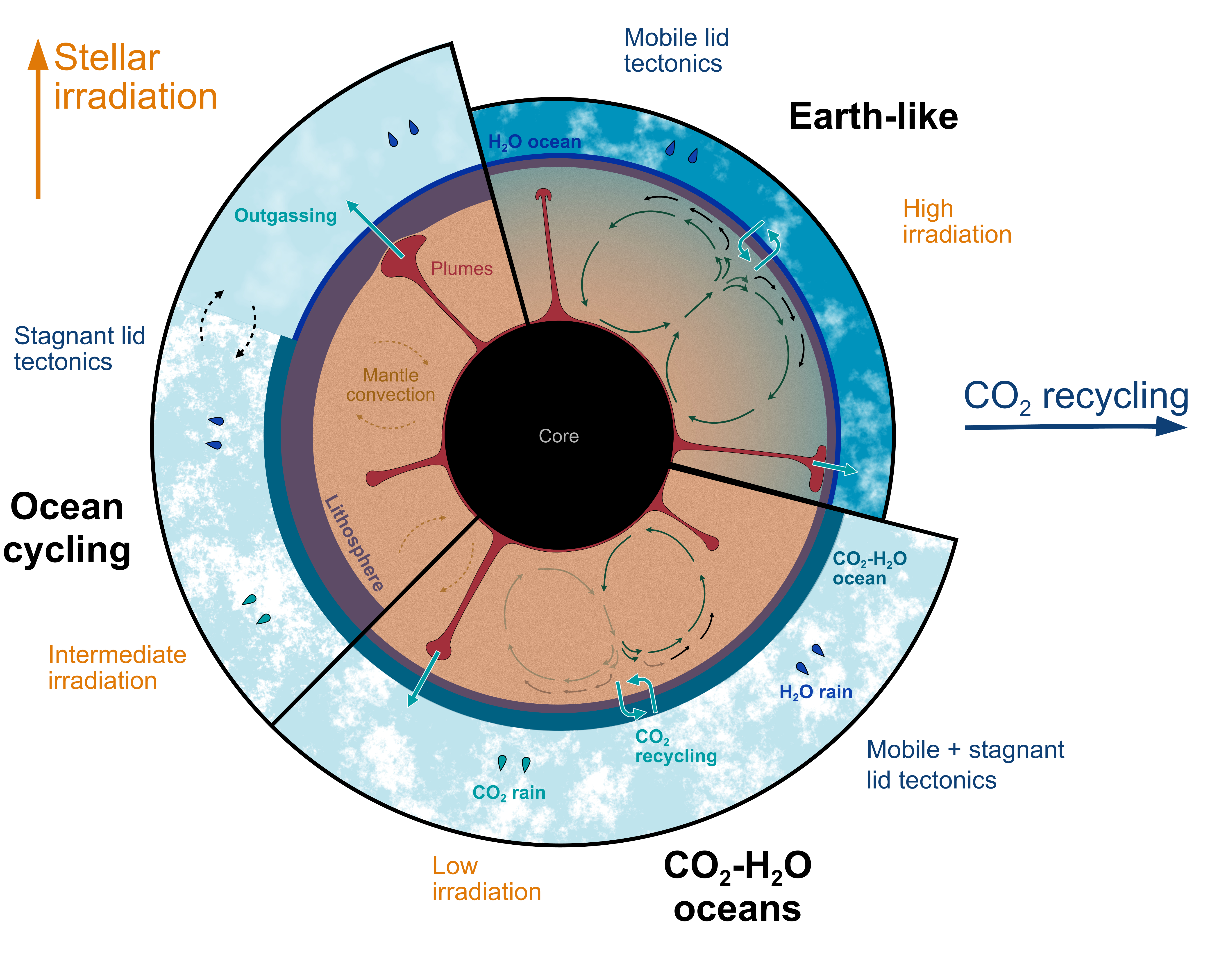}}
    \caption{A schematic illustrating the climate states examined in this article. Planets with high (Earth-like) CO$_2$ recycling driven by efficient weathering and mobile lid tectonics that receive high stellar irradiation are likely to occupy Earth-like climate states with H$_2$O oceans and relatively thin atmospheres. Planets with CO$_2$ recycling that may range from low to high, with either mobile lid or stagnant lid tectonics, and which receive low stellar irradiation, may occupy climate states with thick CO$_2$ atmospheres and surface CO$_2$ condensation, producing oceans with both liquid CO$_2$ and liquid H$_2$O. Planets with low CO$_2$ recycling at intermediate instellations may oscillate between a CO$_2$-H$_2$O ocean state and a state with only H$_2$O oceans. }
    \label{fig:illustration}
\end{figure*}

\subsection{Conclusion}
In summary, our simulations suggest that the interplay of the radiative properties of \ce{CO2}-rich atmospheres and the weathering of silicates leads to super-saturated and cyclic climates for planets under low irradiation for G-type and more massive stars. A qualitative sketch of the distinct climate regimes suggested by our study is shown in Fig.~\ref{fig:illustration}. The cooling impact of Rayleigh scattering at very high \ce{CO2} levels introduces bistability between a climate state where silicate weathering maintains \ce{CO2} at relatively low levels and a state where \ce{CO2} outgassing outpaces silicate weathering, maintaining \ce{CO2} at such high levels that it condenses at the surface. At intermediate instellation, planetary temperatures can become large enough for water vapor in the atmosphere to significantly impact planetary albedo through absorption of incoming light, destabilizing surface \ce{CO2} condensing climate states and giving rise to limit cycles between \ce{CO2} condensing and non-condensing states. The dynamic interplay between radiation and carbon cycling profoundly impacts the climate state and surface geochemistry of otherwise Earth-like planets in the outer reaches of the liquid water habitable zone. These CO$_2$-condensing climate states are potentially distinguishable by observational characterization of CO$_2$ dimer features and a trend in $p$CO$_2$ versus instellation opposite to that anticipated from the nominal carbonate-silicate cycle feedback.

\section*{Open Research}
We have archived the data necessary to reproduce the plots in this article at \citet{graham_bistability_dataset}.
\section*{Acknowledgements} 
RJG acknowledges scholarship funding from the Clarendon Fund and Jesus College, Oxford. TL was supported by a grant from the Simons Foundation (SCOL Award No. 611576). RTP is supported by European Research Council Advanced Grant EXOCONDENSE (Grant No. 740963). This
AEThER publication is also funded in part by the Alfred P. Sloan
Foundation under grant G202114194. We thank Robin Wordsworth, Jim Kasting, and an anonymous reviewer for helpful reviews that significantly strengthened the paper. 

\FloatBarrier
\pagebreak

\appendix
\onecolumn

\begin{table}[htb!]
\centering
\begin{tabular}{lcccc}
Molecule & A & B  & $\Delta$&  $\sigma_0$\\
 & [units of & [units of & & [10$^{-7}$\\ 
  & 10$^{-6}$] & 10$^{-3}$] &  & m$^2$ kg$^{-1}$] \\ \hline
\ce{CO2} & 4.39 & 6.4 &  0.0805 & -- \\
\ce{N2} & 2.906 & 7.7 &  0.0305 & --\\
\ce{H2O} & -- & -- & -- & 9.32\\
\end{tabular}
\caption{Tabulation of Rayleigh scattering data used in this study. A and B values come from \citet{cox2015allen}. $\sigma_0$ for \ce{H2O} is drawn from \citet{Pierrehumbert:2010-book}. $\Delta$ values come from \citet{vardavas1984solar}.}
\label{tab:rayleigh}
\end{table}

\begin{longtable}[htbc]{|p{2.3cm}|p{2.7cm}|p{4.8 cm}|p{2.1cm}|}
\hline
\textbf{Parameter} & \textbf{Units} & \textbf{Definition} & \textbf{Fiducial Value} \\
\hline
$\gamma$&$\textbf{--}$ & Land fraction & 0.3\\
\hline
$a_g$ & $\textbf{--}$ & Surface albedo & 0.0\\
\hline
$R_{\rm planet}$ & meters (m) & Planetary radius & 6.37$\times10^6$\\
\hline
$T_{\rm ref}$& Kelvin (K) & Reference global-& 288\\
&&avg. temperature &\\
\hline
$p$CO$_{\rm 2,ref}$& bar & Reference &280$\times10^{-6}$\\
&&\ce{CO2} partial pressure&\\
\hline
$q_{\rm ref}$& m yr$^{-1}$ & Modern global-&0.20\\
&&avg. runoff&\citep{oki2001global}\\
\hline
$\epsilon$& 1/K &Fractional change in& 0.03\\
&&precip. per K change in temp.&\\
\hline
$V_{\rm ref}$ &mol yr$^{-1}$&Modern global& 7.5$\times 10^{12}$\\
&&\ce{CO2} outgassing &\citep{gerlach2011volcanic,haqq2016limit}\\
\hline
$v$ & mol m$^{-2}$ yr$^{-1}$&Modern \ce{CO2} outgassing&  0.0147\\
&&per m$^2$ planetary area&\\
\hline
$\Lambda$&variable&Thermodynamic coefficient&1.4$\times10^{-3}$\\
&&for $C_{eq}$&\\
\hline
$n$&\textbf{--}&Thermodynamic $p$\ce{CO2}&0.316\\
&&dependence&\\
\hline
$\alpha$* & $\textbf{--}$ & $L\phi\rho_{sf}AX_r\mu$& 3.39$\times 10^5$\\
&&(see Section \ref{subsec:mac_model} and below)&\\
\hline
$k_{\rm eff,ref}$*&mol m$^{-2}$ yr$^{-1}$&Reference rate constant&8.7$\times10^{-6}$\\
\hline
$\beta$ & $\textbf{--}$ & Kinetic weathering & 0.2\\
& &$p$\ce{CO2} dependence &\citep{rimstidt2012systematic}\\
\hline
$T_{\rm e}$ & Kelvin & Kinetic weathering & 11.1\\
& &temperature dependence  &\citep{Berner:1994p3295}\\
\hline

\caption{Model parameters used in this study. This table lists parameters used in our calculations, their units, their definitions, and the default values they take. A single asterisk (*) means the default parameter value was drawn from Table S1 of the supplement to \citet{maher2014hydrologic}. For default parameters drawn from other sources, the citation is given in the ``Value'' column.
\label{tab:weathering_values}
}
\end{longtable}

\FloatBarrier
\pagebreak
\twocolumn
\bibliographystyle{aasjournal.bst}

\bibliography{biblio2,biblio0}



\end{document}